\documentstyle[aps,epsf]{revtex}
\textwidth  
175mm
\begin{document}
\thispagestyle{empty}

\begin{flushright} JLAB-THY-98-10 \\
hep-ph/9803316
\end{flushright}

\begin{center}
{\Large \bf 
Nonforward 
Parton Densities  and}
 \\ \vspace{1mm}{\Large \bf Soft Mechanism for  Form Factors and}
  \\ \vspace{1mm}
{\Large \bf Wide-Angle Compton Scattering in QCD}
\end{center}

\begin{center}
{A.V. RADYUSHKIN}\footnote{Also
Laboratory of Theoretical Physics, JINR, Dubna,
Russian Federation}  \\
{\em Physics Department, Old Dominion University,}
\\{\em Norfolk, VA 23529, USA}
 \\ {\em and} \\
{\em Jefferson Lab,} \\
 {\em Newport News,VA 23606, USA}
\end{center}
\vspace{2cm}

\begin{abstract}

We argue that at moderately large momentum transfer
$-t \lesssim 10 \, {\rm GeV}^2$,
hadronic form factors and wide-angle 
Compton scattering amplitudes are  dominated
by  mechanism corresponding
to overlap of soft wave functions.
We show that   the soft contribution in both cases 
can be described in terms of the same universal 
 nonforward parton densities (ND's) ${\cal F}(x;t)$,
which are the simplest hybrids of
the usual parton densities and hadronic form factors.
We propose a simple model for ND's
possessing required reduction properties.  
Our  model  easily reproduces the observed magnitude
and the dipole $t$-dependence   of the  
proton form factor  $F_1^p(t)$ in 
the region  \mbox{$1\,{\rm GeV}^2 < -t < 10\, {\rm GeV}^2$}.
Our results for the wide-angle Compton scattering 
cross section follow the angular dependence of 
existing data
and   are rather close to the data
in magnitude.

\end{abstract}

\newpage

\section{ Introduction}

The Compton scattering in its various versions 
provides a unique tool for studying 
 hadronic structure.
The  Compton amplitude probes the hadrons through a 
 coupling of two electromagnetic currents
 and in this aspect it  can be considered 
as a generalization of hadronic form factors.  
In QCD, the photons interact with the 
quarks of a hadron through 
a vertex which, in the lowest
 approximation,  has a   pointlike structure.
However, in the soft regime, 
strong interactions  produce large 
corrections uncalculable within the
perturbative QCD framework.
To take advantage of  the basic pointlike structure of the
photon-quark coupling and the asymptotic
freedom feature of QCD, one should choose 
a specific kinematics  in which the behavior
of the relevant amplitude is dominated by 
short (or, being more precise, lightlike) distances.
The general feature of all such types of kinematics is the 
presence of a large momentum transfer.
For  Compton amplitudes, there are several
situations when large momentum transfer induces
dominance of configurations 
involving lightlike distances:  \\ 
$i)$ both photons are far off-shell 
and have  equal spacelike virtuality:
 virtual forward Compton amplitude,
 its imaginary part determines structure
 functions of deep inelastic scattering (DIS); \\ 
 $ii)$ initial photon is highly virtual,
 the final one is real and the momentum transfer 
 to the hadron is small: deeply virtual 
 Compton scattering (DVCS) amplitude; \\ 
 $iii)$ both photons  are real but the 
 momentum transfer
 is large:  wide-angle Compton 
 scattering (WACS) amplitude, the 
 study of which is the ultimate  goal of  the present paper.

 Our main statement 
 is that, at accessible momentum transfers
 $|t| \lesssim 10$ GeV$^2$, the  WACS amplitude  is dominated 
 by  handbag diagrams, just like  in  DIS and DVCS.
 In the most general case, the nonperturbative part of the handbag contribution
 is described by  nonforward double distributions (DD's) 
 $F(x,y;t),   G(x,y;t)$, etc.,  which can be 
 related  to the usual parton 
 densities $f(x)$, $\Delta f(x)$ and nucleon form factors
 like $F_1(t),G_A(t)$. 
 Among the arguments of DD's,   $x$ is 
  the fraction of the initial hadron momentum carried 
 by the active parton and $y$ is the  fraction   
 of the momentum transfer $r$.
The description of WACS amplitude simplifies 
when one can neglect the $y$-dependence 
 of the hard part and integrate out
    the $y$-dependence
 of the double distributions. In that case,
 the long-distance dynamics is described by nonforward
 parton densities (ND's)  ${\cal F}(x;t), {\cal G}(x;t),$ etc. 
 The latter 
 can be interpreted as the usual parton densities $f(x)$ 
 supplemented by a form factor type $t$-dependence.  
 We  propose  a simple model 
 for the relevant ND's 
 which both satisfies the relation between  ${\cal F}(x;t)$ and
   usual parton densities $f(x)$ and  produces 
   a good description of 
  the  $F_1(t)$ form factor up to $t \sim  - 10$ GeV$^2$. 
  We use this model to calculate  the WACS amplitude
  and obtain results which are 
  rather close to existing data.

\section {Virtual Compton amplitudes}

The forward virtual  Compton amplitude whose
imaginary part gives structure functions
of deep inelastic scattering (see, e.g., \cite{feynman})
is the classic example  of a 
light cone dominated Compton amplitude.
 In this case,
the ``final'' photon  has momentum $q'=q$ coinciding with
that of the initial one. The momenta $p,p'$ of the initial
and final hadrons also coincide. 
The total cm energy of the photon-hadron system
$s= (p+q)^2$ should be above resonance region,
and the Bjorken ratio $x_{Bj} = Q^2/2(pq)$ is finite.
The light cone   dominance is secured by high
virtuality of the photons: $-q^2 \equiv Q^2 \gtrsim 1 $ GeV$^2$.
In the large-$Q^2$ limit,  the leading  contribution 
in the lowest $\alpha_s$  order 
is given by  handbag diagrams in which the 
perturbatively calculable hard quark
propagator is convoluted with parton distribution
functions $f_a(x)$ ($a=u,d,s, \ldots$) 
which describe/parametrize nonperturbative
information about hadronic structure.

The condition that both photons are highly virtual
may be relaxed by taking 
a  real photon in the final state.
Keeping the momentum transfer $t\equiv (p-p')^2$ 
to the hadron  as small as possible,
one arrives at  kinematics of the deeply virtual 
Compton scattering (DVCS) the  importance of which 
was recently emphasized by X. Ji \cite{ji}
 (see also \cite{compton}). 
 Having   large  virtuality $Q^2$  
of the initial photon is sufficient to  
guarantee that in the Bjorken limit 
the leading power  contributions in $1/Q^2$ 
are generated  by the strongest  light cone    
singularities \cite{ji2,npd,jios2,cofre},  with   
the handbag diagrams being the  
starting point of the $\alpha_s$
expansion.  
The most important contribution to the 
DVCS amplitude is given  by a convolution
of a  hard quark propagator and 
a nonperturbative function describing 
long-distance dynamics, which in  the most general case is 
given by   nonforward double    distributions 
(DD's) $F(x,y;t), G(x,y;t), \ldots $ \cite{compton,npd}.

\begin{figure}[t]
\mbox{
   \epsfxsize=16cm
 \epsfysize=6cm
  \epsffile{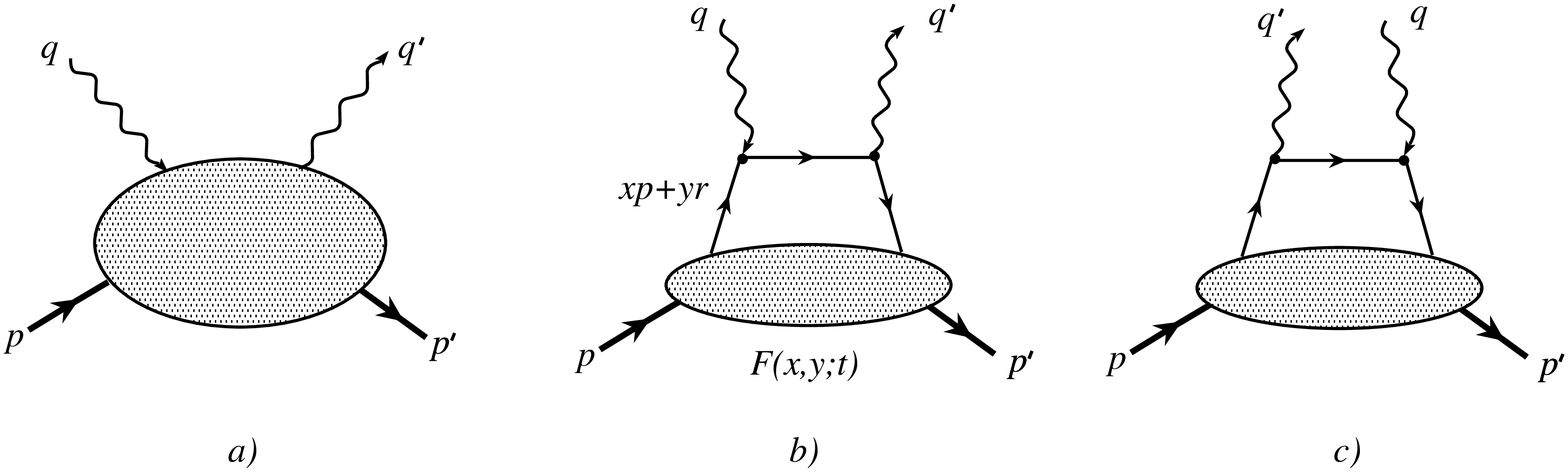}  }
{\caption{\label{fig:cohan} $a)$ General Compton amplitude;
$b)$ $s$-channel handbag diagram;  $c)$ $u$-channel handbag diagram.
   }}
\end{figure}

The DD's are  rather complicated functions.
They specify the fractions $xp$ and $yr$ 
of the initial hadron momentum $p$ and the
momentum transfer $r\equiv p-p'$ carried by the active parton:
$k \sim  x p + yr$. The DD's  vanish outside the 
triangle region 
$0 \leq x+y \leq 1$ \cite{compton,npd}.
In addition to $x$ and $y$,  they also  depend on 
the invariant momentum transfer $t=(p'-p)^2$. 
In  some  limiting cases, 
the double distributions reduce
to simpler and already known functions.
For  $r=0$, 
the matrix elements coincide with  the 
forward ones defining the usual  
parton densities.
This  results in the following ``reduction 
relations'' \cite{compton,npd}
\begin{equation} \int_0^{1-x} F^a(x,y;t=0) \, dy 
= f_a(x)  \ .\label{1}
\end{equation}
Integrating properly weighted sums of quark and antiquark 
DD's  over $x$
one obtains the Dirac   form factor: 
\begin{equation}\sum_a e_a \int_0^1  dx\, \int_0^{1-x}  
 \left [F^a(x,y;t) - F^{\bar a}(x,y;t) \right ] \, dy
 =F_1(t) \, , \label{2}
\end{equation} 
where $e_a$ is the electric charge of the
``$a$'' quark.
Just like for form factors,
one should take into account
 extra double  distributions
$  K^{ a}(x,y;t) $ 
corresponding to a hadron helicity flip in the 
nonforward matrix element \cite{ji}. 
These distributions are related to the Pauli form factor $F_2(t)$:
one should just substitute $F^{a, \bar a}$ 
by $K^{a, \bar a}$ and $F_1$ by $F_2$
in Eq.(\ref{2}).
A  common element of these reduction formulas 
is  an integration over $y$.  Hence, it is convenient 
to introduce   intermediate functions
\begin{equation}
{\cal F}^a(x;t) = \int_0^{1-x}  
 F^a(x,y;t)  \, dy  \ \  ;  \   \  {\cal K}^a(x;t) = \int_0^{1-x}  
 K^a(x,y;t)  \, dy \, .  \label{3a} 
\end{equation}
 They satisfy the reduction formulas
\begin{equation}  {\cal  F}^a(x;t=0) 
= f_a(x)  \  \  ;  \  \   \sum_a e_a \int_0^1 
\left [ {\cal F}^a(x;t) -   
 {\cal F}^{\bar a} (x;t)  \right ] 
 \,  dx  
 =F_1(t) \,  \label{3b} \end{equation}
 \begin{equation} 
\sum_a e_a \int_0^1   
 \left [ {\cal K}^a (x;t) - 
{\cal K}^{\bar a} (x;t)  \right ] 
 \,  dx  =F_2(t) \, , \label{3c}
\end{equation}
which show that these functions are 
the simplest  hybrids of the usual parton densities 
and form factors.
For this reason, we  call them 
 {\it nonforward parton densities} (ND's). 
Note that the $t=0$ limit of the  ``magnetic'' 
ND's  exists:
${\cal K}^a (x;t=0) \equiv k_a(x)$.
In particular, the integral
\begin{equation}
\sum_a e_a \int_0^1   
 \left [ k_a(x) - k_{\bar a}(x) 
  \right ] 
 \,  dx  = \kappa_p \, \label{4}
\end{equation}
gives the anomalous magnetic moment of the proton. 
The  knowledge of the $x$-moment  of  $k_a(x)$'s
is needed to determine the contribution of the
quark orbital angular momentum to the proton spin \cite{ji}.
Since  the  $K$-type DD's  are always accompanied 
by the $r_{\mu} = p_{\mu}-p'_{\mu}$ factor,  they
 are invisible in deep inelastic
scattering and other inclusive processes 
related to strictly forward $r=0$ 
 matrix elements.

There are also parton-helicity 
sensitive double distributions $ G^{ a}(x,y;t)$ and  
$ P^{ a} (x,y;t)$. 
The first one  
reduces to the usual spin-dependent 
densities $\Delta f_a(x)$ in the 
$r=0$ limit and gives the axial 
form factor $F_A(t)$ after the $x,y$-integration.
The second one involves hadron helicity flip
and is  related to the pseudoscalar form factor 
 $F_P(t)$.

In the DVCS kinematics,   $|t|$ is assumed to be small compared to 
$Q^2$, and for this reason 
the $t$- and $m_p^2$-dependence of the short-distance amplitude 
in refs. \cite{ji,compton,ji2,npd} was neglected\footnote{
One should not think that  such a dependence 
is necessarily a higher
twist effect: the lowest twist contribution
has a calculable dependence on $t$ and $m_p^2$ 
analogous to the Nachtmann-Georgi-Politzer 
 $O(m_p^2/Q^2)$ target mass corrections in DIS
\cite{nachtmann,geopol}.}. 
This is equivalent to approximating the 
active parton momentum $k$ by its plus component alone:
$k \to  xp^+ + yr^+$. Treating  $\zeta \equiv  r^+ / p^+$
as an  external parameter  and  using  the 
total fraction $X \equiv x+ \zeta y$ as an
independent variable,
one arrives at an alternative description 
of the DVCS  scaling limit in terms of the 
nonforward parton distributions 
\cite{gluon,npd} (NFPD's)\footnote{Other 
terminology: ``off-forward'' \cite{ji}, 
``non-diagonal'' \cite{cfs} 
and  ``off-diagonal'' \cite{diehl,maryskin} is also used in the literature.
Off-forward parton distributions introduced 
by X. Ji  in his 
pioneering papers on  DVCS \cite{ji,ji2} are 
equivalent though not identical to
 the nonforward ones,
while ``non-diagonal''
and  ``off-diagonal'' distributions 
essentially coincide with  NFPD's,  see \cite{npd} 
for details.}  ${\cal F}_{\zeta}(X;t)$.
They are related to  double distributions  by
\begin{equation}
{\cal F}_{\zeta}^{a,\bar a}(X;t) =
 \int_0^{\min \{X/\zeta, \bar  X / \bar \zeta \}}
F^{a,\bar a} (X- \zeta y,y;t) \, dy \, .
\label{4b}
\end{equation} 
In a similar way, one can incorporate  the relevant 
double distributions to define  also 
``magnetic'' ${\cal K}^{ a, \bar a}_{\zeta}(X;t)$ 
and parton-helicity 
sensitive nonforward distributions 
${\cal G}^{ a, \bar a}_{\zeta}(X;t)$ and  
${\cal P}^{ a, \bar a}_{\zeta}(X;t)$ \cite{ji,brad,bgr}.
In addition to the usual parton  momentum fraction variable $X$ 
and the invariant momentum transfer $t$,
the NFPD's   also depend on 
the skewedness parameter $\zeta = r^+/p^+$ 
 specifying the longitudinal
momentum asymmetry of the nonforward matrix element.
This asymmetry appears because it is impossible 
to convert a highly virtual initial photon into a real
final photon without a longitudinal momentum transfer.
In general, one can use different pairs of vectors
to specifiy the longitudinal direction:  
$(p,q)$,   $(p,q')$ or $(P,q)$ with $P=(p+p')/2$, etc.,
resulting in different $t$-dependent expressions for $\zeta$.
However, in the (formal) scaling 
limit $t \to 0, m_p^2 \to 0$
all these expressions  for the 
skewedness parameter $\zeta$ coincide with 
the Bjorken ratio $x_{Bj}=Q^2/2(pq)$ \cite{compton,npd}.

\section{Modeling ND's}

Our final  goal in the present paper is to  get an estimate  of 
  the handbag  contributions 
for the large-$t$ real Compton scattering.
 Since the initial   photon in that case is also     real: 
 $Q^2=0$ (and hence $x_{Bj}=0$), it is natural to expect that 
 the nonperturbative functions which appear in WACS 
 correspond   to the $\zeta = 0$ limit of 
 the  nonforward parton distributions\footnote{Provided 
 that one can neglect the $t$-dependence of the hard 
 part, see a footnote above and discussion in Sec. VI.} 
 ${\cal F}^{ a}_{\zeta}(x;t)$.
 It is easy to see from Eqs.(\ref{3a}),(\ref{4b}) 
 that in this limit
 the NFPD's reduce to the  nonforward parton 
densities ${\cal F}^a(x;t)$
introduced above:  
\begin{equation} {\cal F}_{\zeta=0}^a(x;t) = 
{\cal F}^a(x;t) \, .\end{equation}
 Note that ND's depend on  
 ``only   two''  variables $x$ and $t$,
 with this dependence  
   constrained by reduction
 formulas (\ref{3b}),(\ref{3c}).  
 Furthermore,    it is possible 
to give an interpretation of nonforward  densities  
in terms of the  light-cone wave functions.

Consider for simplicity 
a two-body bound state  whose 
lowest Fock component is described by a light cone   
wave function $\Psi(x,k_{\perp})$.
Choosing a frame where the momentum transfer $r$  is purely 
transverse  $r=  r_{\perp}$, we can write  
the two-body contribution into the form factor 
as \cite{bhl}
\begin{equation}
F^{(tb)} (t) = \int_0^1 \,  dx \, \int \, 
\Psi^* ( x, k_{\perp}+\bar x r_{\perp}) \,  
\Psi (x, k_{\perp}) 
\, {{d^2 k_{\perp}}\over{16 \pi^3}}  \, , 
\end{equation}  
\begin{figure}[htb]
\mbox{
   \epsfxsize=12cm
 \epsfysize=5cm
 \hspace{3cm}  
  \epsffile{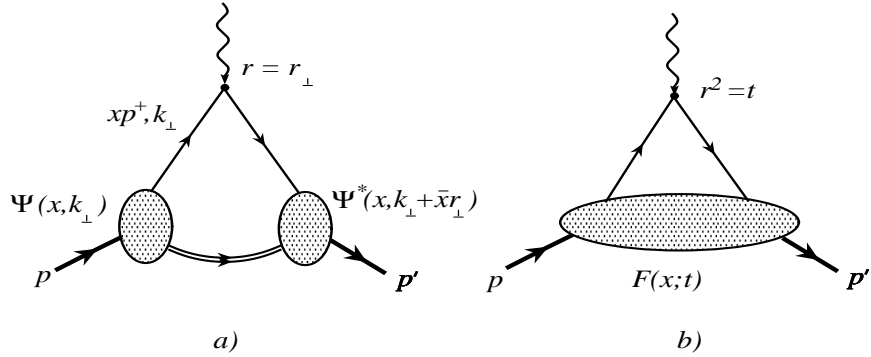}  }
{\caption{\label{fig:psifor} $a)$ Structure of the 
effective two-body contribution 
to form factor in the light cone formalism.
$b)$ Form factor as an $x$-integral  of  nonforward parton densities.
   }}
\end{figure}
\noindent where $\bar x \equiv 1-x$. 
  Comparing this expression with the reduction formula 
  (\ref{3b}), we conclude that 
 \begin{equation} 
 {\cal F}^{(tb)} (x,t) = \int  \, 
 \Psi^* ( x, k_{\perp}+ \bar x r_{\perp}) \,
\Psi (x, k_{\perp}) \, 
{{d^2 k_{\perp}}\over{16 \pi^3}}
\end{equation}  
 is the  two-body contribution into the nonforward  parton density
 ${\cal F} (x,t)$.
Assuming  a Gaussian dependence on the transverse momentum $k_{\perp}$
(cf. \cite{bhl})    
\begin{equation}\Psi (x,k_{\perp}) =  \Phi(x) 
 e^{-k^2_{\perp}/2x \bar x \lambda^2} \,  , \label{11} 
 \end{equation}
we get 
\begin{equation}
{\cal F}^{(tb)} (x,t) = f^{(tb)}(x) e^{\bar x t /4 x \lambda^2 } \, , \label{8}
\end{equation}
where 
\begin{equation}
f^{(tb)}(x) = 
\frac{x \bar x  \lambda^2}{16 \pi^2} \, \Phi^2(x) 
= {\cal F}^{(tb)} (x,t=0)
\end{equation}
is the two-body part of the relevant parton density.
Within the light-cone approach, to get the total
result for either  usual $f(x)$
or 
nonforward parton densities   ${\cal F}(x,t)$,
one should 
add the contributions due to  higher Fock components.
By no means these contributions
are small, e.g., the  valence $\bar d u$  contribution
into the normalization of the $\pi^+$  form factor 
at $t=0$ is less than 25\% \cite{bhl}. 
In the absence of a formalism providing  explicit expressions
for an infinite tower of light-cone wave functions 
 we choose to treat  Eq.(\ref{8}) as a guide
for  fixing interplay between the $t$ and $x$ dependences
of ND's and propose to 
 model them by 
\begin{equation}
{\cal F}^a(x,t) = f_a(x) e^{\bar x t /4 x \lambda^2 }  = 
{{f_a(x)}\over{\pi x \bar x  \lambda^2}}\,
\int  \, e^{-(k^2_{\perp}+ (k_{\perp}+
\bar x r_{\perp})^2)/2x \bar x \lambda^2}
d^2 k_{\perp} \,  . \label{13}
\end{equation}
The functions  $f_a(x)$  here are the 
usual parton densities assumed to be 
taken from existing  parametrizations like GRV, MRS, CTEQ, etc.
In the $t=0$
limit (recall that $t$
is negative)  this model, by construction,  
satisfies the  first of  reduction formulas (\ref{3b}). 
 Within the Gaussian ansatz (\ref{13}), 
 the basic scale $\lambda$ specifies  the 
  average transverse momentum carried by the quarks.
   In particular, for valence quarks 
  \begin{equation} \langle  k^2_{\perp} \rangle ^a = 
  \frac{\lambda^2}{N_a}\int_0^1 
  x \bar x f_a^{val}(x)  \,  dx   \, , 
  \end{equation}
  where $N_u=2, N_d=1$ are the numbers of 
  the valence $a$-quarks in the
  proton.

  To fix the magnitude of  $\lambda$, we  
  use the second reduction formula in 
  (\ref{3b}) relating ${\cal F}^a(x,t)$'s
  to the $F_1(t)$ form factor.
  To this end, we take the following simple expressions for
  the valence distributions
  \begin{equation} f_u^{val} (x) = 1.89 \,
   x^{-0.4} (1-x)^{3.5} (1+6x) \, , \end{equation}
  \begin{equation} f_d^{val} (x) = 0.54 \, 
   x^{-0.6} (1-x)^{4.2} (1+8x) \, . \end{equation}
   \begin{figure}[htb]
\mbox{
   \epsfxsize=8cm
 \epsfysize=8cm
 \hspace{4cm}  
  \epsffile{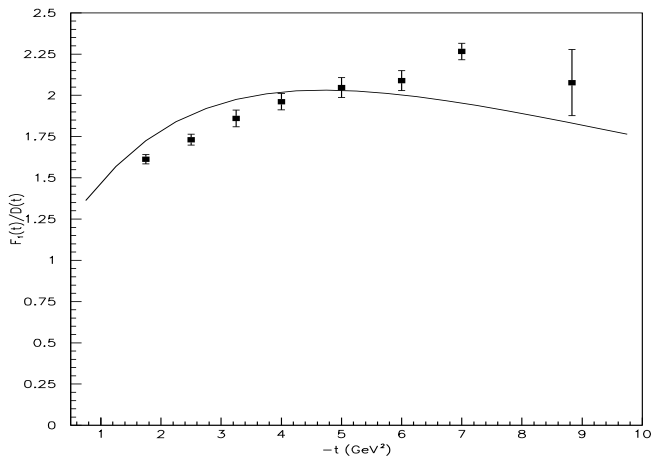}  }
{\caption{\label{fig:ff} Ratio $F_1^p(t)/D(t)$ 
of the $F_1^p(t)$ form factor to
the dipole fit $D(t) =1/(1-t/0.71\,{\rm GeV^2})^2$. Curve
is based on  Eqs.(16)-(18) with $\lambda ^2 = 0.7 \,
{\rm GeV}^2$. Experimental
data are taken from ref.[18].
   }} 
\end{figure} 
\noindent They  closely reproduce the  relevant 
 curves given by the GRV parametrization \cite{grv} 
 at a low normalization point $Q^2 \sim 1$  GeV$^2$. 
 The best agreement between our  model 
 \begin{equation}
 F_1^{\rm soft}(t) =  \int_0^1 \left [ e_u\, f_u^{val} (x) +e_d
  \, f_d^{val} (x) \right ] e^{\bar x t / 4x \lambda^2} dx \label{14}
  \end{equation}
  and experimental data \cite{ff} in the
  moderately large $t$ region
   \mbox{1  GeV$^2$ $< |t|< 10$ GeV$^2$} is reached for 
   $\lambda^2 =0.7 \,$ GeV$^2$ (see Fig.\ref{fig:ff}). 
   This value gives a reasonable magnitude 
   \begin{equation}  \langle  k^2_{\perp} \rangle^u = (290 \,  {\rm MeV})^2 
   \  \   \  ,  \ \ \   \langle  k^2_{\perp} \rangle^d = (250 \, {\rm MeV})^2
   \end{equation}
   for the average transverse momentum of the valence $u$ and $d$ quarks
   in the proton.

 Similarly, building a model  for the 
parton helicity sensitive ND's ${\cal G}^a(x,t) $
one can take their  $t=0$ shape  from existing 
parametrizations for spin-dependent 
parton distributions $\Delta f_a(x)$
and  then fix the relevant $\lambda$ parameter by fitting
the $G_A(t)$ form factor. 
The case of hadron spin-flip distributions ${\cal K}^a(x,t)$ 
and ${\cal P}^a(x;t)$ is more complicated  
since the distributions $k_a(x)$, $p_a(x)$ are unknown.

At $t=0$, our model by construction gives a  correct 
normalization $F_1^p(t=0)=1$ for the form factor.
However, if one would try to find the derivative
$(d/dt)F_1^p(t)$ at $t=0$ by expanding the exponential 
$\exp [\bar x t/x \lambda^2]$ into the Taylor series under 
the integral (\ref{14}), one would get a divergent 
expression.  An analogous  problem is well known 
in applications of QCD sum rules to form factors 
at small $t$ \cite{iosmil,balyung,nestradsmall,belkogan}.
The divergence is related to the  
long-distance propagation of massless quarks in the $t$-channel.
Formally, this is revealed  by singularities  starting at $t=0$. 
However, $F_1^p(t)$ should not have singularities
for timelike $t$ up to $4 m_{\pi}^2$, with the $\rho$-meson 
peak at $t = m_{\rho}^2 \sim 0.6 \,{\rm GeV}^2$
being the most prominent feature of the $t$-channel spectrum.
Technically, the singularities of the original
expression are singled out into the  bilocal correlators
\cite{bal} which are substituted by  their 
realistic version with  correct spectral properties
(usually the simplest model with $\rho$ and $\rho '$
terms is used).  An important point
is that  such a modification is needed only 
when one calculates form factors in the  small-$t$ region:
for  $-t\gtrsim 1\,{\rm GeV}^2$,  the  correction terms
should vanish  faster than any power of $1/t$ \cite{nestradsmall}. 
In our case, the maximum deviation of the curve for $F_1^p(t)$ given     
by  Eq.(\ref{14}) from the experimental data  in the small-$t$ region  
$-t\lesssim 1\, {\rm GeV}^2$  is 15$\%$. 
Hence, if one is willing  to tolerate such an inaccuracy,
one can use our model starting with $t=0$.

\section{Soft vs hard contributions to  form factors}

Our curve  is within 5\% from the data  points \cite{ff} for
$1 \, {\rm GeV}^2 \lesssim -t \lesssim 6$ \, GeV$^2$ and does not deviate 
from them by more than 10\% up to 9 GeV$^2$.
Modeling the $t$-dependence by a more complicated formula
(e.g., assuming  a slower decrease at large $t$, and/or
choosing different $\lambda$'s for
$u$ and $d$ quarks and/or splitting ND's
into several components with different 
$\lambda$'s,  etc.) or changing the shape 
of parton densities $f_a(x)$ 
one can improve the quality of the fit
and extend  agreement with the data to higher $t$. 
Such a fine-tuning  is not our goal here. 
We just want to emphasize that 
a reasonable  description of the  $F_1(t)$ data in a wide region 
\mbox{1  GeV$^2$ $< |t|< 10$ GeV$^2$ } 
was obtained by fixing just a single parameter
 $\lambda$ reflecting the proton size. Moreover, we could 
fix $\lambda$ from the requirement that
$\langle  k^2_{\perp} \rangle \sim (300 \,  {\rm MeV})^2$
and present   our curve for $F_1(t)$ as a successful prediction 
of  the  model.    
We interpret this success as  an 
 evidence that the model
correctly catches the gross features of the 
underlying physics.

Since 
our   model implies  a Gaussian  
dependence on the transverse momentum,
it includes  only what is usually referred to as   
an overlap of soft wave functions.
It completely  neglects effects due to 
hard pQCD gluon exchanges  generating 
the power-law  $O( (\alpha_s / \pi)^2 /t^2)$
tail of the nonforward densities at large $t$.
It is worth pointing out here that though we take    nonforward 
densities  ${\cal F}^a(x,t)$ with  an exponential 
dependence on $t$, 
the $F_1(t)$ form factor in our model has 
a power-law asymptotics    $F_1^{\rm soft} (t) \sim (-4 
\lambda^2/t)^{n+1}$ 
dictated  by the   $(1-x)^n$ behavior of 
the parton densities for $x$ close to 1.
This connection arises because  
the integral (\ref{14}) over $x$ 
is dominated at large $t$ by  the region 
$\bar x \sim  4 \lambda^2 /|t|$. 
In other words,  the large-$t$ behavior of $F_1 (t) $ in our model
is governed by the Feynman mechanism \cite{feynman}. 
One should realize, however, that  the relevant scale
$4 \lambda^2 =2.8$ GeV$^2$ is rather large.
For this reason, when 
 $|t| < 10$ GeV$^2$,  
 it is premature to rely on asymptotic estimates 
 for the soft contribution. 
  Indeed, with $n=3.5$, the asymptotic estimate is  
 $F_1^{\rm soft} (t) \sim t^{-4.5}$,
in an apparent contradiction with 
the ability of our curve to follow 
the  dipole behavior.
The resolution of this paradox is very simple:
the maxima of nonforward distributions  ${\cal F}^a(x,t)$ 
for $|t| \lesssim 10$ GeV$^2$ are 
at rather low $x$-values $x \lesssim 0.5$.
Hence, the $x$-integrals producing $F_1^{\rm soft} (t)$
are not dominated  by the $x \sim 1$ region yet and the 
asymptotic  estimates are not applicable:
the functional dependence of $F_1^{\rm soft} (t) $ in 
our model  is much more complicated than 
a simple power of $1/t$.    
 
  The fact that our    model  closely reproduces 
the experimentally 
observed 
dipole-like behavior of the proton form factor is a 
clear demonstration 
  that 
such a  behavior may have nothing to do with   the
quark counting rules $F_1^p(t) \sim 1/t^2$ \cite{brofar,mmt}
valid  for the
asymptotic behavior of 
the  hard gluon exchange 
contributions.
 Our explanation of the observed 
 magnitude and the $t$-dependence
 of  $F_1  (t)$ by a purely soft 
 contribution  is in  strong contrast   
 with that of the hard pQCD  approach to this problem. 
Of course, there is no doubt that 
 in the formal  asymptotic limit $|t|  \to \infty$,
the dominant contribution to the $F_1(t)$ 
form  factor in QCD  is given
by diagrams involving two hard gluon exchanges,
with nonperturbative dynamics described
by distribution amplitudes (DA's) $\varphi_p(x_1,x_2,x_3)$,  
$\varphi_p(y_1,y_2,y_3)$ of the initial and final
protons \cite{blnuc,cz}. 
However,  attempting to  describe  the data 
at accessible $t$ by   hard contributions only, 
  one is forced to make several unrealistic assumptions.
  
  The crucial element is the use   of humpy DA's
   similar to those 
 proposed by 
Chernyak  and I. Zhitnitsky \cite{cznuc,cz} (CZ).  
The usual claim is that these DA's are 
backed by QCD sum rule calculations of their lowest 
moments. However, as we argued in ref. \cite{mikhrad},
a straightforward version of the  QCD sum rule approach  in this case is  
unreliable because of poor convergence 
of the underlying operator product expansion (OPE).
In the  analysis of the QCD sum rules
for the moments of the pion 
distribution amplitude performed in refs. \cite{mikhrad,nonloc},
the contribution  of exploding higher terms of the OPE 
(neglected in the CZ approach) 
was modeled by  nonlocal condensates.  
The resulting   QCD sum rule produces  the pion DA close to the 
asymptotic one. 
The statement that the pion DA is close 
to its asymptotic form even at a low normalization point
is also supported    by 
a lattice calculation of the second moment 
of the pion DA \cite{gupta}, by QCD sum rule 
estimate of the magnitude of $\varphi_{\pi}(x)$ 
at the middlepoint $x=1/2$ \cite{brfil}, 
by the analysis of QCD sum rules for the 
nondiagonal correlator \cite{minn,mikhbak},  
by calculation of the
pion DA in the chiral soliton model \cite{petpob} 
and by a direct QCD sum rule calculation of the 
large-$Q^2$ behavior of the 
$\gamma^* \gamma \pi^0$ form factor \cite{rrnp}.
Furthermore, within the  light-cone QCD sum rule approach 
one can relate the pion DA to the pion parton densities \cite{beljo}
known experimentally.
According to the analysis performed in \cite{beljo2},
existing data  favor the  asymptotic shape. 
Finally, 
the humpy  pion DA  advocated  in 
\cite{czpi,cz}  
is now ruled out by recent experimental data \cite{cleo} 
on the $\gamma^* \gamma \pi^0$ form factor. 
The data  are fully consistent with the next-to-leading 
pQCD prediction  calculated using  the asymptotic DA
\cite{braaten,murad,brojipar}.

Since  the structure of OPE in the pion and 
nucleon cases is  very similar,  
we see no reason   
to expect  a significant  deviation of the nucleon DA  
from its asymptotic form. 
In particular, an evidence against humpy nucleon 
DA's is provided by a lattice calculation \cite{marsac} 
which does not indicate any significant asymmetry. 
One may argue 
that the proton DA  must be asymmetric to  
reflect  the  fact that
the $u$-quarks carry on average a larger  fraction 
of the proton momentum than the $d$-quarks. 
As shown in ref. \cite{bokroll},   to 
accomodate this observation one needs only 
a moderate shift of the DA maximum from the center point
$x_1=x_2=x_3=1/3$.
 Such a shift does not produce a  drastic enhancement
 of the hard contribution provided by the humpy  DA's.
 However, with the asymptotic DA,
the leading twist hard contribution  completely 
fails to describe the data: 
 it gives   zero for the proton magnetic form factor
and a wrong-sign (positive)  contribution for the neutron
magnetic form factor,
with the  absolute magnitude of the latter being two orders 
of magnitude below the data \cite{belioffe}.

 Furthermore, as emphasized  in refs. \cite{ils,dourdan},
   the  whole strategy of getting enhancements  
 from the humpy  DA's is  based on  an implicit 
 assumption  that one 
 may use the  perturbative expressions 
 $S(k) \sim \hat k /k^2$, $D(k) \sim 1/k^2$ for quark 
 and gluon propagators up to very small virtualities 
 $k^2  \lesssim (300 \, {\rm MeV})^2$. 
It is worth recalling  now    why  CZ-type DA's
 give an enhanced contribution.
  Since  quarks in the proton 
 carry only a fraction  of the
 proton momentum, the characteristic 
 virtualities  $\sim x_i y_j t$ of ``hard'' quarks and gluons inside 
 the short-distance subprocess are   smaller than 
 the total momentum transfer $t$. 
 For a symmetric distribution, one would expect that
 $\langle x_i \rangle \sim 1/3$. 
 With  the humpy  DA's,   the average $x_i$ for one of the 
 $u$-quarks is close to 1, and the dominant contribution
 comes from configurations in which   this quark is active.
 Then   fractions $x_i$ related 
 to passive quarks are rather small. 
 It is precisely the small magnitude of the $\sim x_i y_j t$
 denominators of quark and gluon 
 propagators which produces the enhancement in the case of
  the  CZ-type DA's. 
Hence, to get  large hard contributions,
it is absolutely necessary to assume   
that the  perturbative expressions 
 $S(k) \sim \hat k /k^2$, $D(k) \sim 1/k^2$ for quark 
 and gluon propagators  may be trusted  
 up to very small virtualities.

 An instructive illustration of possible modifications
 due to finite size or transverse momentum effects 
 is given by the light-cone calculation 
 of the $\gamma^* \gamma \pi^0$ 
 amplitude \cite{bhl,murad} in which hard  
 propagator of a {\it massless}  
 quark is convoluted with the two-body wave function
 $\Psi (x,k_{\perp})$.
 Assuming the   Gaussian dependence \mbox{$\Psi (x,k_{\perp}) \sim
 \exp[ - k_{\perp}^2/2 x \bar x \sigma]$} on  transverse momentum,
 one can easily calculate the $k_{\perp}$ integral to see
   that the pQCD propagator factor $1/xQ^2$ is substituted
 by the combination $(1- \exp[ - xQ^2/2  \bar x \sigma])/xQ^2$
 which  monotonically 
 tends to a finite limit $1/2 \sigma$ as $x \to 0$.
 Hence, the effective virtuality is always larger than $2 \sigma$. 
 The suppression
 of propagators at low virtualities has a simple 
 explanation: propagation of quarks and gluons in the 
 transverse direction 
 is restricted by the finite size of the hadron. 
Numerically, 
$2 \sigma \approx 1.35 \, {\rm GeV}^2$ in that case.   
 However, even a milder  modification of  the ``hard''  
 propagators by 
  effective quark and gluon 
 masses $1/k^2 \to 1/(k^2 - M^2)$ with 
 $M^2 \sim 0.1 \, {\rm GeV}^2$ or model 
  inclusion of transverse momentum effects
 strongly reduces the magnitude of  hard contributions
 \cite{jakroll}, especially
 when the CZ type DA's are used.
   For these reasons,  a  scenario 
 with humpy DA's and bare $\sim 1/x_iy_jt$ propagators  
 (which amounts to ignoring finite-size effects)   
considerably overestimates   the size  of hard contributions. 
   
   The  relative smallness of  hard contributions
   can be easily understood within the QCD sum rule 
   context. The soft contribution  is dual to the
   lowest-order diagram while the gluon exchange 
   terms appear in diagrams having a higher order 
   in $\alpha_s$ which results in the usual 
   $\alpha_s/\pi \sim 1/10$  suppression factor 
   per each extra loop. 
   In particular, the $\alpha_s/\pi $ suppression factor is 
   clearly visible in the expression for the hard contribution
   to the pion form factor \cite{czs,farjack,er,bl} 
\begin{equation}    
   F_{\pi}^{\rm hard}(Q^2) |_{\varphi_{\pi} = \varphi_{\pi}^{as}}
   = \frac{8 \pi \alpha_s f_{\pi}^2}{Q^2} = 2 
   \left ( \frac{\alpha_s}{ \pi} \right  )  \frac{s_0}{Q^2} \, .
  \end{equation} 
Here, the combination $s_0 = 4 \pi^2 f_{\pi}^2 
\approx 0.67$ GeV$^2 \sim m_{\rho}^2 $ 
is what is usually called the  ``typical  
hadronic  scale'' in the case of the pion.   
   At asymptotically high
   $Q^2$, the  $O(\alpha_s/\pi)$
    suppression of the hard terms is more than compensated
   by their slower decrease with $Q^2$. 
   However, such a compensation does not occur
   in the subasymptotic region where  
   the soft contributions, as we have seen, may 
    have the same effective  power 
   behavior as that  predicted by the asymptotic
   quark counting rules for the hard contributions.
    In  ref. \cite{sjir},  both  the soft contribution
   and the $O(\alpha_s)$ corrections for the pion form factor 
   were calculated together within a QCD sum rule
   inspired  approach.
   The ratio of the $O(\alpha_s)$ terms to the 
   soft contribution  was shown to be in full agreement 
   with the expectation based on the 
    $\alpha_s/\pi $ per loop suppression.

\section{ Compton scattering amplitude 
at large momentum transfer}

With both photons real, it is not sufficient to have
large photon energy to ensure short-distance dominance:
large-$s$, small-$t$ region is strongly affected by
Regge contributions.
Hence, having large $|t| \gtrsim 1 \, $GeV$^2$ is a  necessary
condition for revealing   short-distance  dynamics.

\begin{figure}[htb]
\mbox{
   \epsfxsize=10cm
 \epsfysize=4cm
  \epsffile{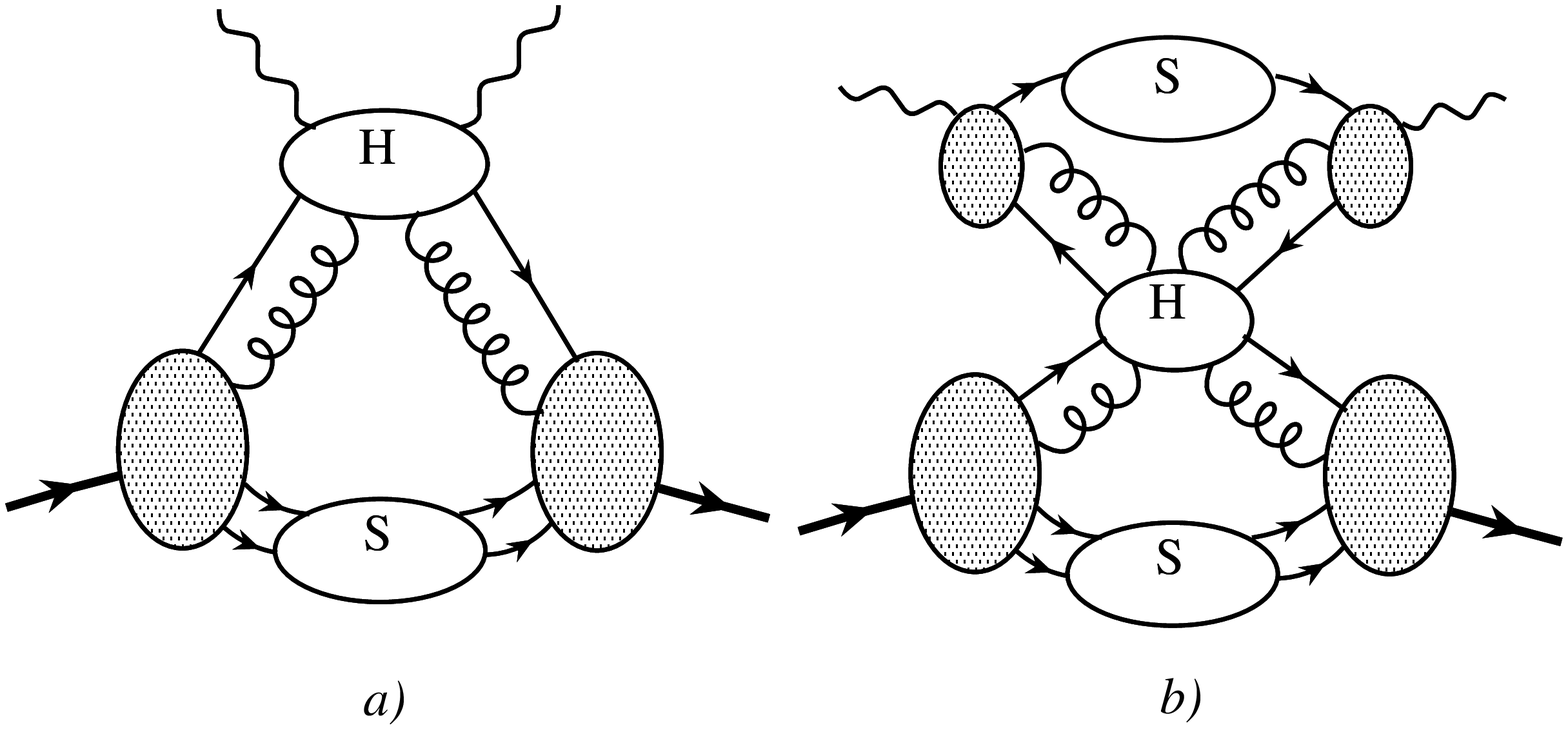} \hspace{-1.3cm} \epsfxsize=8cm
 \epsfysize=4cm \epsffile{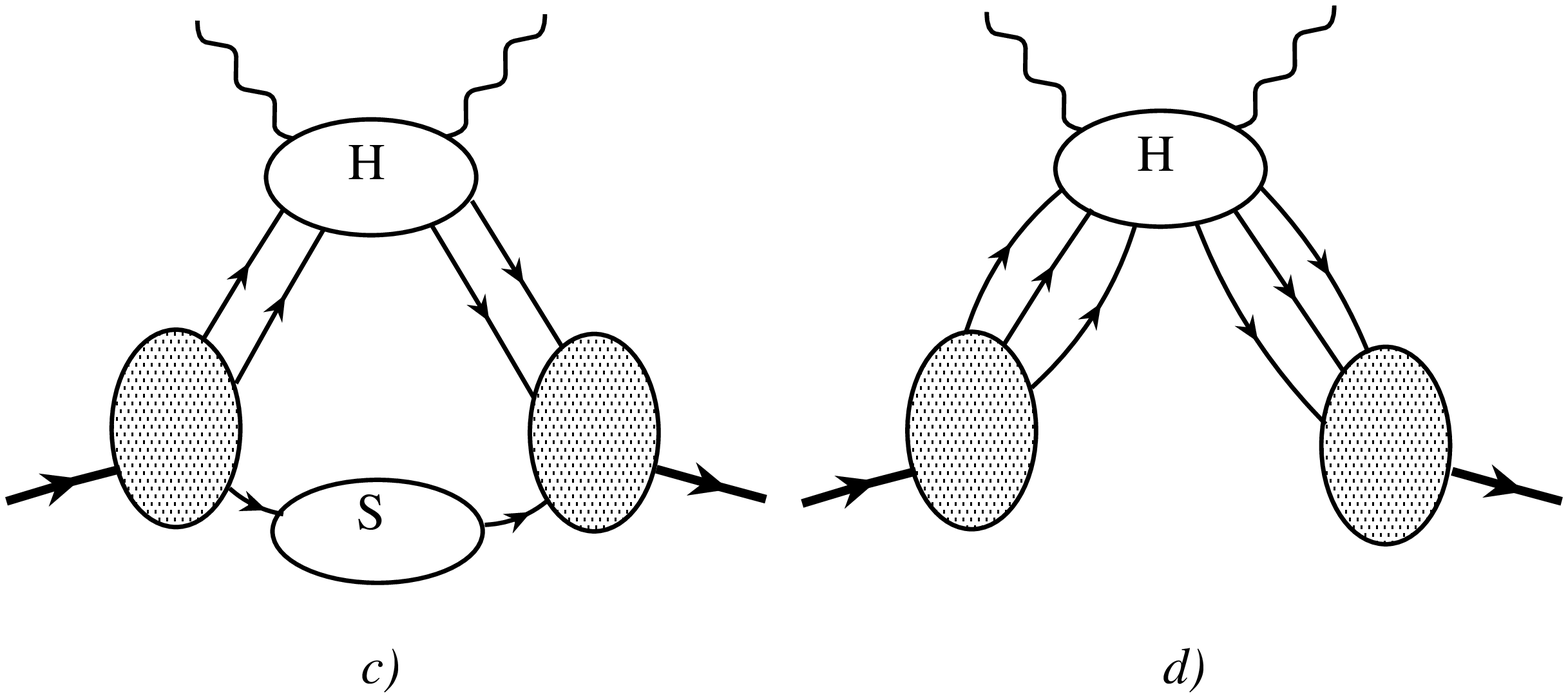} } 
{\caption{\label{fig:cofac} Some configurations responsible
for power-law asymptotic contributions for the WACS amplitude.
   }}
\end{figure}

Consider  the Compton scattering 
amplitude  for   large 
values of  $s$, $u$ and  $t$.   
 According to a general rule (see, e.g., \cite{npd}
 and references therein),  to find 
 possible mechanisms generating  power-law
 contributions in the  asymptotic limit 
 $s \sim -u \sim -t \sim Q^2$ ($Q$ here is just
 a characteristic scale),
 we should look for subgraphs whose 
 contraction into point  or removal from the diagram  
 kills its  dependence 
 on large variables.  
 Contracted  subgraphs correspond to short-distance (SD)
 or hard   
 regime while  the removed  ones to the infrared (IR) or soft  regime.
  Some possibilities   are shown 
 in Fig.\ref{fig:cofac}.

 The power counting estimate for each SD-subgraph
 is given by 
 \begin{equation}
 A_H(Q) \sim Q^{4-N-\Sigma{_i} t_i} \, ,
\end{equation}
where $N$ is the number of the external photon lines
of the hard  subgraph $H$  and $t_i$ is the twist
of  its $i$th external parton line ($t=1$ for quarks and
physical gluons and $t=0$ for longitudinal gluons).

The perturbative estimate for an  IR contribution 
is given by 
 \begin{equation}
 A_S(Q) \lesssim Q^{-\Sigma{_j} t_j} \, ,
\end{equation}
where summation is over the external lines 
of the soft subgraph $S$. 
The infrared regime corresponds to the Feynman 
mechanism. However, we should keep in mind that the  
perturbative estimate implies a pointlike 
coupling of three quarks to the proton field
while in real life the proton 
wave function is much softer. 
In particular, the perturbative estimate of   the 
IR regime for the proton 
form factor gives  $F(Q^2) \lesssim Q^{-4}$,
allowing for $1/Q^4$ behavior in principle. 
To get such an asymptotic  behavior from our ND models,
we should assume that $f(x) \sim 1-x $   
for $x$ close to 1. More realistic functions 
dictate a  faster decrease of $F(Q^2)$  in the asymptotic $Q \to \infty$
limit. No wonder:  the IR regime is essentially 
 nonperturbative and the  $Q$-dependence 
 of the  soft  contributions should be better 
 taken from a  reasonable 
 model rather than from perturbation theory.
Again, since  
accessible $Q$'s are far from being asymptotic,
the ``observed'' power behavior of the soft contribution
in this region may strongly differ from the asymptotic 
powers given by the Feynman mechanism.

\begin{figure}[htb]
\mbox{
   \epsfxsize=16cm
 \epsfysize=4cm
  \epsffile{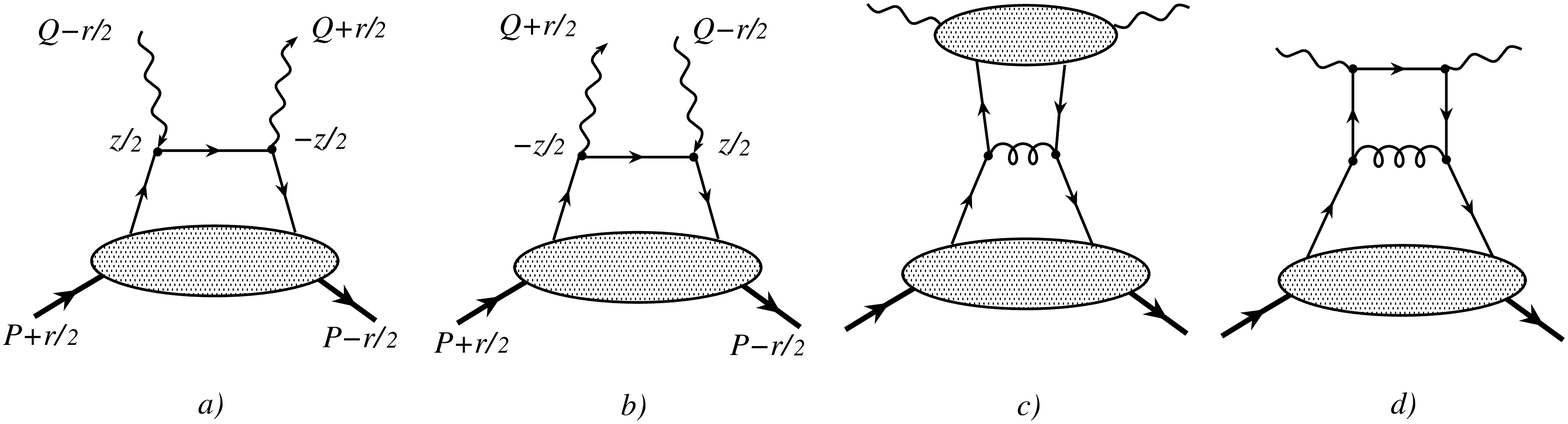}   } 
  \vspace{0.5cm}
{\caption{\label{fig:cosuble} Terms having  $O(s^0)$ 
behavior for large $s$.
   }}
\end{figure}

The simplest contributions for the WACS amplitude 
are given by the $s$- and $u$-channel
handbag diagrams Fig.\ref{fig:cosuble}a,b. They 
correspond to a combined SD-IR regime of Fig.\ref{fig:cofac}a: 
the  dependence on $s$ (or $u$) is killed
by contracting into point the quark line
connecting the photon vertices 
while the $t$-dependence is killed by removal
of a soft subgraph $S$. 
The SD-regime in this case gives $A_H(s) \sim s^0$ behavior.
The nonperturbative part is given by the proton 
nonforward DD's  
which determine the $t$-dependence of the total contribution.
Another  $O(s^0)$ configuration is shown in Fig.\ref{fig:cosuble}c.
In this case, a hard gluon propagator is convoluted with
the proton and photon DD's. Similarly to the usual photon 
structure functions, the photon DD's can be divided 
into the perturbative and the nonperturbative part.
The latter  corresponds to hadronic component of the real
photon while the first one to a direct pointlike 
quark-photon coupling. It can be treated as a part
of the one-loop correction  to the handbag diagram
(see Fig.\ref{fig:cosuble}d) and is accompanied by the 
$\alpha_s/\pi $ suppression factor.
The hadronic component of the photon DD's 
has also an extra form factor type suppression
$\sim m^2/t$.

Just like in the form factor case, 
the contribution dominating in the 
formal asymptotic limit $s,|t|, |u| \to \infty$,
is given
by diagrams corresponding to the pure SD regime,
see Fig.\ref{fig:cofac}d. The hard subraph $H$ 
involves two hard gluon exchanges which results in  
suppression by a factor $(\alpha_s/ \pi)^2 \sim 1/100$  
absent in  the handbag term. 
The total contribution  of all 
hard configurations  was calculated by Farrar 
and Zhang
\cite{far} and then recalculated by  
 Kronfeld and Ni\v{z}i\'{c} \cite{kroniz}. 
 Again,   a sufficiently large 
 contribution is  only obtained if one uses  humpy DA's
 and   $1/k^2$ propagators with no finite-size effects included.
 Even with such propagators, 
 the WACS  amplitude 
calculated with the asymptotic DA is 
negligibly small \cite{vander} compared to existing  data. 
Our arguments concerning the reliability of  CZ enhancements
for form factors can be applied to the 
wide-angle Compton scattering without any changes. 
For these reasons, we ignore  hard contributions 
to the WACS  amplitude as negligibly small.

\begin{figure}[htb]
\mbox{
   \epsfxsize=16cm
 \epsfysize=4cm
  \epsffile{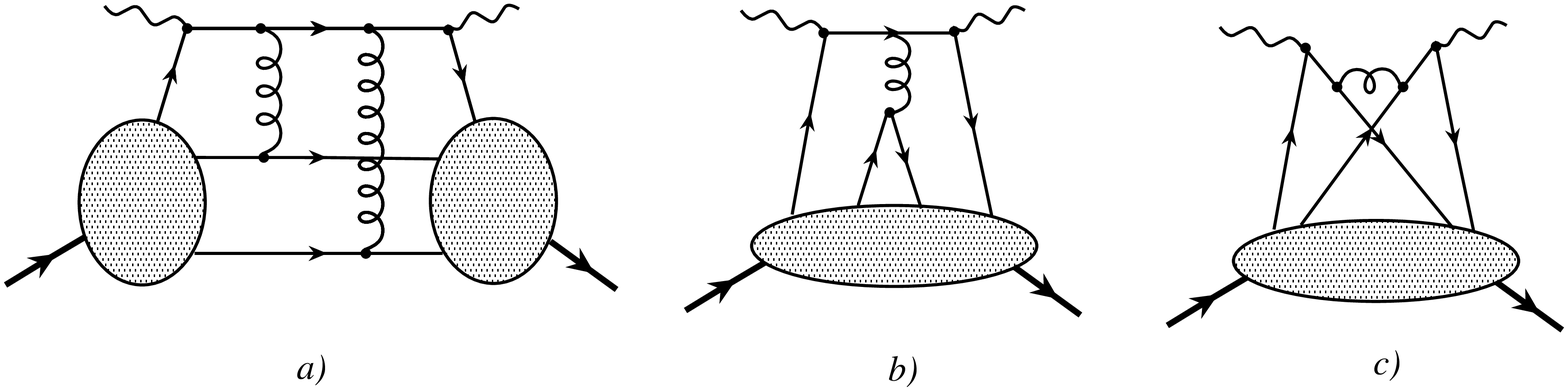}   }   
  \vspace{0.5cm}
{\caption{\label{fig:cosubno} Configurations 
involving double and single gluon exchange.
   }}
\end{figure}

Another type of configurations containing  
 hard gluon exchange corresponds to the version of the 
combined SD-IR regime shown in Fig.\ref{fig:cofac}c.   
 In particular, they include diagrams like Fig.\ref{fig:cosubno}b 
 and also diagrams with 
 photons coupled to different 
quarks (``cat's ears'', Fig.\ref{fig:cosubno}c). 
Such contributions have   both higher order 
and   higher twist.
This brings in the 
$\alpha_s/\pi $ factor  and  
 an  extra  $1/s$ suppression.
The latter is  partially compensated by  a slower fall-off 
of the four-quark DD's with $t$ since only one valence quark
should change its momentum.

\section{Model for wide-angle Compton scattering amplitude}

\begin{figure}[htb]
\mbox{
   \epsfxsize=7cm
 \epsfysize=6cm
  \epsffile{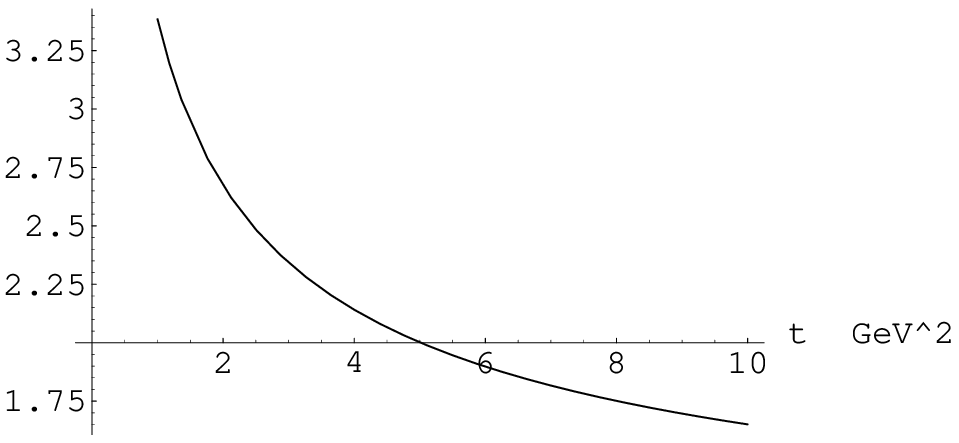} \hspace{2cm}    \epsfxsize=7cm
 \epsfysize=6cm
  \epsffile{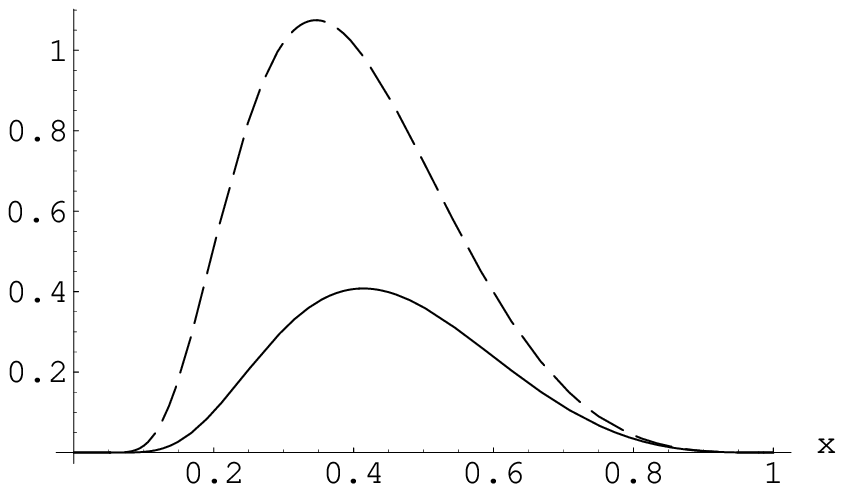} }
  \vspace{0.5cm}
{\caption{\label{fig:rcuux} $a)$ Ratio
$R_1^u(t)/F_1^u(t)$; $b)$ Functions  
${\cal F}^u (x;t)$ (solid line)
 and ${\cal F}^u (x;t)/x$ (dashed line) at  $t = - 2.5$ GeV$^2$. 
  }}
\end{figure}

In this paper, we neglect all the suppressed terms 
and deal only with the handbag contributions Fig.\ref{fig:cosuble}a,b
in which 
the highly virtual quark propagator connecting the photon vertices 
  is convoluted with nonforward proton DD's
  parametrizing the 
overlap of soft wave functions.  
Since  the  basic scale $4 \lambda^2$ 
characterizing the $t$-dependence of 
DD's in our model  is 2.8 GeV$^2$, while existing data    are 
all  at  momentum transfers $t$ below   5  GeV$^2$,
we deal with the region where the asymptotic estimate 
(Feynman mechanism) for the overlap
contribution is  not working yet.
In the coordinate representation, the sum of two 
handbag contributions to 
the Compton amplitude  can be written as 
\begin{equation}
M^{\mu \nu} (p,p';q,q') = \sum_a e_a^2 \int  e^{-i(Qz)} 
\langle p'|
 ( \bar \psi_a (z/2) \gamma^{\mu} S^c(z) \gamma^{\nu} \psi_a (-z/2)
+ \bar \psi_a (-z/2) \gamma^{\nu} S^c(-z) 
\gamma^{\mu} \psi_a (z/2)) |p  \rangle \, 
d^4z
\end{equation}
where $Q = (q+q')/2$ and
$S^c(z) = i \hat z/2  \pi^2 (z^2)^2 $ 
is the hard quark propagator 
(throughout, we use the 
``hat'' notation $\hat z \equiv z_{\alpha}\gamma^{\alpha}$).
The summation over the twist-0 longitudinal gluons
adds the usual gauge link between the $\bar \psi$,$ \psi$ fields
which we do not write down explicitly 
(gauge link disappears, e.g.,  in  the Fock-Schwinger gauge 
$z^{\alpha} A_{\alpha} (z) =0$).
Because of the symmetry of the problem, it is convenient to use 
$P = (p+p')/2$ (cf. \cite{ji}) and $r = p-p'$ as the basic 
momenta. 
Applying the  Fiertz transformation and  introducing the double  
distributions by 
\begin{eqnarray}
\langle p' |  \bar \psi_a (-z/2) \hat z
\psi_a (z/2) |p \rangle \, = 
\bar u(p') \hat z u(p) \, \int_0^1 \, dx \int_{-\bar x/2}^{\bar x/2} 
 \left [ e^{-i(kz)} {\tilde F}^a(x,\tilde y;t) 
- e^{i(kz)} {\tilde  F}^{\bar a} (x,\tilde y;t) \right ] \, d\tilde y \nonumber \\ 
+ \frac1{4m_p}  \, \bar u(p') (\hat z \hat r - \hat r \hat z)  u(p)
\int_0^1 \, dx \int_{-\bar x/2}^{\bar x/2} \, \left [e^{-i(kz)} {\tilde K}^a(x,\tilde y;t) 
- e^{i(kz)} {\tilde  K}^{\bar a} (x,\tilde y;t) \right ] 
\, d\tilde y  + O(z^2) \  {\rm terms}
 \label{17} \end{eqnarray}
(we use here   the  shorthand notation $k 
\equiv xP+ \tilde yr$) and similarly for 
the  parton helicity sensitive operators
\begin{eqnarray} 
\langle \, p'  \,  | \,   \bar \psi_a(-z/2) \hat z \gamma_5
 \psi_a(z/2) 
\, | \,p  \,  \rangle \,  
=    \bar u(p')  \hat z  \gamma_5
 u(p) \,  \int_0^1 \, dx \int_{-\bar x/2}^{\bar x/2} 
 \left [  e^{-i(kz)}
 {\tilde  G}^a(x,\tilde y;t)  +  e^{i(kz)}
{\tilde  G}^{\bar a}(x, \tilde y;t)  \right ] 
 \,  d\tilde y \nonumber \\
+
\, \frac{(rz)}{m_p} \,  \bar u(p')  \gamma_5
 u(p) \,    \int_0^1 \, dx \int_{-\bar x/2}^{\bar x/2} 
  \left [ e^{-i(kz)}
 {\tilde  P}^a(x,\tilde y;t)  +  e^{i(kz)}
{ \tilde P}^{\bar a}(x,\tilde y;t)  \right ] 
 \,  d\tilde y + O(z^2) \  {\rm terms} \,  ,  \label{18}
 \end{eqnarray} 
we arrive at  a leading-twist QCD parton picture 
with tilded DD's  serving as functions 
describing long-distance dynamics. 
The new  DD's ${\tilde F}^a(x, \tilde y;t)$,  
 etc., are related to the original 
 DD's ${ F}^a(x,y;t)$ discussed   
 in Section II by the shift $y= \tilde y + \bar x/2$.
 Integrating  ${\tilde F}(x, \tilde y;t)$ over  
 $\tilde y$ one obtains the same  
 nonforward densities  ${\cal F}(x ;t)$.
  The hard quark propagators  
for the $s$ and $u$ channel handbag diagrams 
in this picture  look like
 \begin{equation}
\frac{x \hat P + \tilde y \hat r + \hat Q}{(xP+\tilde yr+Q)^2} = 
\frac{ x \hat P + \tilde y \hat r+ \hat Q}{x \tilde s -
(\bar x ^2 /4 -\tilde y^2)t +x^2 m_p^2} 
\  \  {\rm and } \  \ 
\frac{x \hat P + \tilde y \hat r - \hat Q}{(xP+\tilde yr-Q)^2} =
\frac{x  \hat P+ \tilde y \hat r-\hat Q}{x \tilde u -
(\bar x ^2 /4 -\tilde y^2)t +x^2 m_p^2} \, , 
\end{equation}
respectively. We denote $\tilde s =2(pq)=s-m^2$ and  
 $\tilde u =-2(pq')=u-m^2$.
 Since DD's are even functions of $\tilde y$
 \cite{lech}, the $\tilde y \hat r$ terms 
 in the numerators can be dropped. 
It is legitimate to
keep $O(m_p^2)$ and $O(t)$ terms in the denominators: 
 the dependence of  hard propagators 
on target parameters $m_p^2$ and $t$ can be 
calculated exactly because of  the effect analogous to the 
$\xi$-scaling 
in DIS \cite{geopol} (see also \cite{rrnp}).
Note that the $t$-correction to hard propagators disappears 
in the large-$t$  limit dominated by the $x \sim 1$ integration.
The $t$-corrections are the largest for $y=0$. 
At this value and   for $x=1/2$  and $t=u$ (cm angle of 90$^{\circ}$), 
the $t$-term in the denominator of the most important
second propagator is 
only 1/8 of the $u$ term. This ratio increases to 1/3 for 
$x=1/3$. However, at nonzero $\tilde y$-values, the $t$-corrections 
are smaller. Hence, the $t$-corrections in the denominators
of hard propagators can produce $10\% -20 \%$ effects  
and should be included in a complete analysis.     
In the present paper,  
we will consider a simplified approximation in which 
 these terms are neglected 
and  hard propagators are given 
by $\tilde y$-independent 
expressions $(x \hat P + \hat Q)/x \tilde s$ 
and $(x \hat P + \hat Q)/x \tilde u$.
As a result, the  $\tilde y$-integration acts only on the
  DD's ${\tilde F}(x,\tilde y;t)$ and 
converts them into  nonforward densities ${\cal F}(x,t)$. 
The latter    
would  appear then  through two types of integrals
 \begin{equation}
\int_0^1 
{\cal F}^a(x,t) \, {dx} \equiv F_1^a(t) \ \ {\rm and} \ \ \int_0^1 
{\cal F}^a(x,t) \, \frac{dx}{x} \equiv   R_1^a(t),  
\end{equation}
 and similarly for ${\cal K,G,P}$.
 The functions $F_1^a(t)$ are the flavor components 
 of the usual $F_1(t)$ form factor while $R_1^a(t)$
 are  the flavor components of a new form factor
  specific to the wide-angle Compton scattering.
  In the formal asymptotic limit $|t| \to \infty$, the $x$ integrals 
  for $F_1^a(t)$ and $R_1^a(t)$ 
  are both dominated in our model 
  by the $x \sim 1$ region: the large-$t$ 
  behavior  of these functions is governed 
  by the Feynman mechanism and their ratio tends to 1 as
  $|t|$  increases (see Fig.\ref{fig:rcuux}a). However, due to 
  large value of the effective scale $4 \lambda^2 =2.8$ GeV$^2$, 
  the  accessible momentum transfers
  $t \lesssim 5$  GeV$^2$ are very far from being asymptotic.

In Fig.\ref{fig:rcuux}b we plot  
   ${\cal F}^u (x;t)$ and ${\cal F}^u (x;t)/x$ at  $t = - 2.5$ GeV$^2$. 
  It is clear that the relevant integrals  are dominated
  by rather small $x$ values $x \lesssim 0.4$ 
  which results in a strong
  enhancement of $R_1^u(t)$ 
   compared to $F_1^u(t)$ for $|t| \lesssim 5$  GeV$^2$.
   Note also that the 
   $\langle p' | \ldots x \hat P \ldots |p \rangle $ 
   matrix elements  can produce only  $t$ as a large variable 
   while $\langle p' | \ldots  \hat Q \ldots |p \rangle $ 
   gives $s$. As a result, the enhanced form factors 
   $R_1^a(t)$ are accompanied by extra $s/t$ enhancement factors
   compared to the $F_1^a(t)$ terms. In the cross section,
   these enhancements are squared, i.e.,  
   the contributions due to the non-enhanced form factors   $F_1^a(t)$ 
  are always accompanied by $t^2/s^2$ factors
  which are smaller than 1/4 
  for cm angles 
  below 90$^{\circ}$. Because of double suppression,
    we neglect $F_1^a(t)$ 
  terms  in the present simplified 
  approach.

 The  contribution due to the ${\cal K}$ functions  
  appears through  the flavor components $F_2^a(t)$ of the $F_2(t)$
  form factor and their enhanced analogues $R_2^a(t)$. 
  The major part of contributions due to  the ${\cal K}$-type ND's 
  appears in the  combination 
  \mbox{$R_1^2(t)-(t/4m_p^2)R_2^2(t)$.}
  Experimentally, $F_2(t)/F_1(t)\approx 1 \,{\rm GeV}^2/|t|$.
  Since $R_2/F_2 \sim R_1/F_1 \sim 1/ \langle x  \rangle $, 
  $R_2(t)$ is similarly suppressed compared 
  to $R_1(t)$,  and we  neglect contributions 
  due to the $R_2^a(t)$ form factors.  
  We also neglect here the terms with 
  another spin-flip distribution  ${\cal P}$  related
  to the pseudoscalar form factor $G_P(t)$ which is dominated 
  by the $t$-channel pion exchange.  Our  calculations
  show that the contribution due to  
  the parton helicity sensitive densities  ${\cal G}^a$ 
  is suppressed by the factor $t^2/2s^2$ compared to that due to the 
  ${\cal F}^a$ densities. This factor only reaches
  1/8  for the cm angle of  
  90$^{\circ}$,  and hence the ${\cal G}^a$ contributions are not 
  very significant  numerically. For simplicity, we 
  approximate  ${\cal G}^a(x,t) $ by ${\cal F}^a(x,t)$. 
After these approximations,
the WACS  cross section is given by the product  
 \begin{equation}
 \frac{ d \sigma}{dt}  \approx \frac{2 \pi \alpha^2}{\tilde s^2} 
 \left [  \frac{(pq)}{(pq')} +  \frac{(pq')}{(pq)}
  \right ] \,  R_1^2(t) \,  , 
 \end{equation}
 of  the  Klein-Nishina  cross section 
(in which we dropped  $O(m^2)$ and $O(m^4)$ terms) and 
 the square of the $R_1(t)$ form factor 
  \begin{equation} 
 R_1(t) =  \sum_a e_a^2  \left [R_1^a (t) + R_1^{\bar a} (t) \right ] \, .
 \end{equation}
 In our model, $R_1(t)$ is given by 
 \begin{equation} 
R_1(t) =   \int_0^1 
\biggl [ e_u^2 \, f_u^{val} (x) +
  e_d^2  \, f_d^{val} (x) + 2( e_u^2+e_d^2+e_s^2) 
  \, f^{sea} (x) \biggr ] 
  e^{\bar x t / 4x \lambda^2} \frac{dx}{x} \, . 
\end{equation}
 We included here  the 
 sea   distributions assuming that they are all
  equal $f^{sea} (x)
=f^{sea}_{u,d,s} (x)= f_{\bar u, \bar d,\bar s} (x)  $ 
and using  a  simplified parametrization 
\begin{equation}
f^{sea} (x) = 0.5 \,  x^{-0.75} (1-x)^7
\end{equation}
which  accurately reproduces  
 the  GRV formula for $Q^2 \sim 1$ GeV$^2$.
 Due to  suppression  of the small-$x$ region
 by the exponential $\exp [\bar x t / 4x \lambda^2]$,
 the sea quark contribution is rather 
 small ($\sim 10 \%$) even for $-t \sim  1$   GeV$^2$ and 
 is invisible  for  $-t \gtrsim 3 $   GeV$^2$.

\begin{figure}[htb]
\mbox{
   \epsfxsize=8cm
 \epsfysize=11cm
 \hspace{4cm}  
  \epsffile{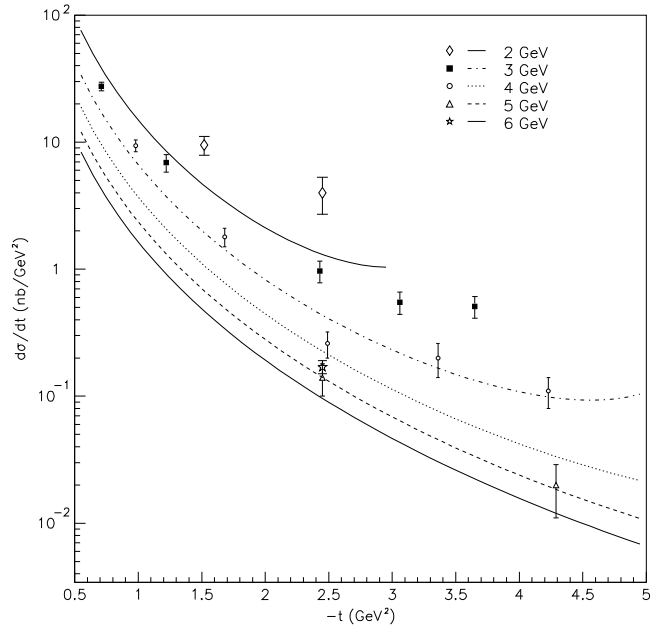}  }
  \vspace{-0.8cm}
{\caption{\label{fig:rct} WACS cross section versus $t$: 
comparison of results 
based on Eq.(28) with experimental data.
   }}
\end{figure}

Comparison with existing data \cite{schupe} 
is shown in Fig.\ref{fig:rct}.  Our curves  follow the data pattern
but are systematically lower  by a factor of  2,
with disagreement becoming more pronounced 
 as  the scattering angle  increases.
Since we neglected several terms each capable 
of producing up to a $20 \%$ correction in  the amplitude, we consider
the agreement between our curves and the data 
as encouraging. The most important corrections which should 
be included in a more detailed investigation
are the $t$-corrections in the denominators of 
hard propagators and contributions due to the ``non-leading''
${\cal K, G,P}$ nonforward densities.
The latter, as noted above, are usually accompanied 
by $t/s$ and $t/u$ factors, i.e., their contribution 
becomes  more significant at larger angles. 
The $t$-correction in the most important hard propagator term  
$1/[x \tilde u - (\bar x^2/4 - \tilde y^2)t +x^2 m_p^2]$
also enhances the amplitude at large angles.

\begin{figure}[htb]
\mbox{
   \epsfxsize=8cm
 \epsfysize=11cm
 \hspace{4cm}  
  \epsffile{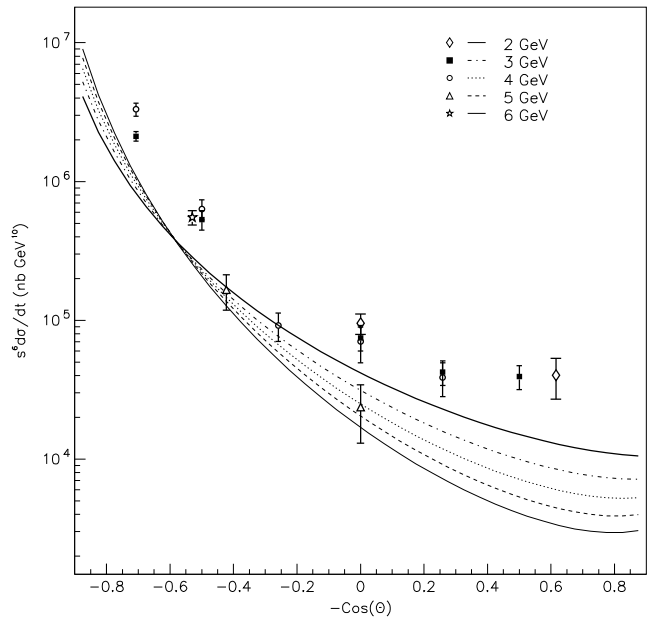}  }
  \vspace{-0.5cm}
{\caption{\label{fig:rctheta}  Angular dependence of 
the combination $s^6 (d \sigma /dt)$.
   }}
\end{figure}

The angular dependence of our results for the combination
$s^6 (d \sigma /dt)$ is shown on Fig.\ref{fig:rctheta}.
All the curves for initial photon ehergies 2,3,4,5 and 6 GeV
intersect each other at  $\theta_{\rm cm} \sim 60^{\circ}$.
This  is in good agreement with experimental data
of ref.\cite{schupe} where the differential 
cross section at fixed cm angles was fitted by powers of $s$:
$d \sigma /dt \sim s^{-n (\theta)}$ with 
$n^{\rm exp}(60^{\circ}) = 5.9 \pm 0.3$. 
Our curves correspond to $n^{\rm soft}(60^{\circ}) \approx 6.1$
and $n^{\rm soft}(90^{\circ}) \approx 6.7$ which also agrees 
with the experimental result $n^{\rm exp}(90^{\circ}) = 7.1 \pm 0.4$.

This can be compared with the scaling behavior
of the asymptotic  hard contribution: 
modulo logarithms contained in the $\alpha_s$ factors,
they have    a universal angle-independent power
$n^{\rm hard}  (\theta) =6$.
For $\theta_{\rm cm}  = 105^{\circ}$, the experimental result
based on just two  data points is $n^{\rm exp}(105^{\circ}) = 6.2 \pm 1.4$,
while our model gives $n^{\rm soft}(105^{\circ}) \approx 7.0$.
Clearly, better data are needed to draw any conclusions here.

\begin{figure}[htb]
\mbox{
   \epsfxsize=8cm
 \epsfysize=11cm
 \hspace{4cm}  
  \epsffile{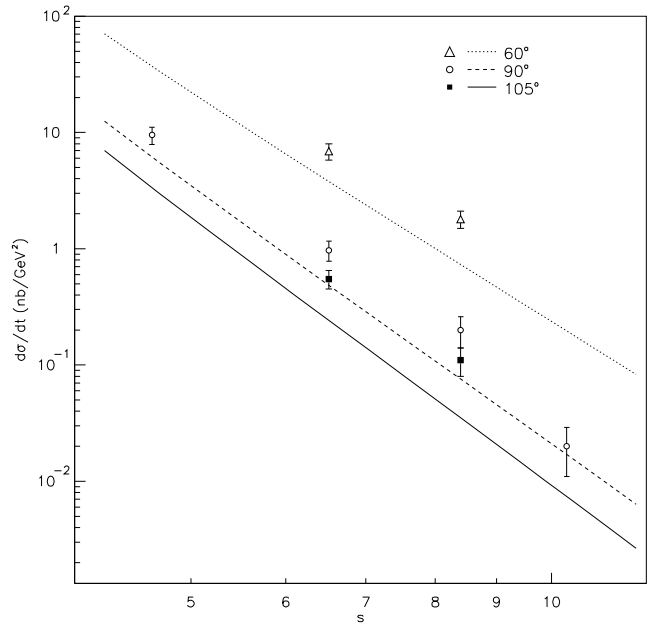}  }
  \vspace{-0.5cm}
{\caption{\label{fig:rcs} $s$-dependence of 
the combination  $s^6 d \sigma /dt $ 
 for $\theta = 60^{\circ}$
(dotted line),  
$\theta = 90^{\circ}$ (dashed line)
and $\theta = 105^{\circ}$ (solid line).
   }}
\end{figure}

 \section{Summary and conclusions} 
 
 In this paper, we  introduced  nonforward 
 parton densities ${\cal F}(x;t)$ which are the simplest
 hybrids of the usual parton densities and hadronic form 
 factors. We   proposed a simple model for 
 the quark ND's ${\cal F}^{a}(x;t)$ which, in the $t \to 0$ limit,  
 reproduces the standard parametrizations for the
 usual parton densities and gives a reasonable
 description of 
 existing  data on the $F_1^p(t)$ form factor
 in a wide range 1 GeV$^2 \lesssim -t  \lesssim $ 10 GeV$^2$ 
 of momentum transfer.  The crucial  observation  
 is that though our model includes only
 the soft contribution,
 the form factor  is dominated 
 at accessible energies by rather small 
 momentum fractions $x \sim 0.5$ and asymptotic
 estimates for  soft contributions 
 (corresponding to Feynman mechanism, i.e.,  dominance 
 of the $x \sim 1$ region) 
 are not working yet. 
 We gave arguments  that the wide-angle Compton  scattering 
 amplitude in the same $t$ region is dominated 
 by two handbag diagrams. We found also that the largest term
 contains the
 same ND's ${\cal F}(x;t)$ which determine the behavior 
 of $F_1^p(t)$.  However, 
 due to the extra $1/x$ factor and small value 
 of $\langle x \rangle$,  
 the WACS amplitude gets a strong enhancement bringing
 our predictions close to existing experimental data.
  Still, there remains a systematic  difference 
  by a factor of 2 between our results and existing data.
   
   On the experimental side,   data of higher quality are 
   needed. They are expected from  a  
    future experiment at 
 Jefferson Lab  \cite{bogdan}, in which better statistical
 accuracy is aimed and several
new ideas will be used to control the systematic errors. 
 
  On the theoretical side, a more detailed approach 
  is needed which would take into account  all nonforward densities.
  A more complete analysis should also include calculable
  $t$- and $m_p^2$ dependence of the hard quark propagators
  and terms which are not enhanced by the
  $1/ \langle x \rangle$ factors.  
  It should be emphasized that 
  keeping   the $t$-terms in the denominators 
 of hard propagators requires a major change
 in the whole approach: it would be  no longer possible 
 to get a simplified description in terms of the 
 nonforward densities ${\cal F}(x;t)$: one should deal 
 then with double distributions ${\tilde F}(x, \tilde y;t)$
 in all their complexity and construct a model 
 for their profile in the $\tilde y$ direction. 
 This observation also demonstrates that
 the double distributions ${ F}(x,  y;t)$ are the 
 primary objects for  analysing   nonforward
 matrix elements of  light cone operators.
 They are more fundamental than their
 reductions such as nonforward, off-forward, etc.
 distributions which work only when the 
 hard part of the relevant amplitude depends 
 on a particular linear combination 
 $x + y \zeta$ of its two arguments $x$ and $y$.
 A more detailed discussion of double distributions
 will be given in a forthcoming publication
 \cite{ddee}.

 \section{Acknowledgement}
 
 This investigation was strongly influenced 
  by  the real Compton scattering 
 enthusiasts C.E. Hyde-Wright, A. Nathan and 
 B. Wojtsekhowski, to whom I  am  most grateful 
 for  numerous discussions. 
I also benefited from discussions 
 and communications with A. Afanasev, 
 I. Balitsky, S. Brodsky, C. Coriano, N. Isgur and 
 I. Musatov. My special gratitude is 
 to I. Musatov for help  with  figures
 and for patiently teaching me how to use
 his {\it Feynman Graph} program \cite{FG}.
 This work was supported by the US 
 Department of Energy under contract
DE-AC05-84ER40150.

\end{document}